\documentclass[preprint]{JHEP}

\usepackage{epsfig}
\usepackage{tabularx}
\usepackage{graphicx}
\usepackage{amssymb,amsfonts,amsmath}

\title{Gravitino dark matter in brane-world cosmology}

\author{Grigoris Panotopoulos\\
ASC, Department of Physics LMU,\\
Theresienstr. 37, 80333 Munich, Germany\\
{\tt E-mail: grigoris@theorie.physik.uni-muenchen.de}}

\abstract{The gravitino dark matter hypothesis in the brane cosmology is studied. The theoretical framework is the CMSSM for particle physics and RS II brane model for gravity. It is found that the gravitino can play the role of dark matter in the universe and we determine what the gravitino mass should be for different values of the five-dimensional Planck mass. An upper bound is obtained for the latter.}

\begin{document}

\maketitle 

\setcounter{equation}{0}

\section{Introduction}

There are good theoretical reasons for which particle physics proposes that new exotic particles must exist. Supersymmetry (SUSY) is an ingredient that appears in many theories for physics beyond the standard model. SUSY solves the hierarchy problem and predicts that every particle we know should be escorted by its superpartner. In order for the supersymmetric solution of the hierarchy problem to work, it is necessary that the SUSY becomes manifest at relatively low energies, less than a few $TeV$, and therefore the required superpartners must have masses below this scale (for supersymmetry and supergravity see e.g. \cite{nilles}).

One of the theoretical problems in modern cosmology is to understand the nature of cold dark matter in the universe. There are good reasons, both observational and theoretical, to suspect that a fraction of $0.22$ of the energy density in the universe is in some unknown ``dark'' form. Many lines of reasoning suggest that the dark matter consists of some new, as yet undiscovered, massive particle which experiences neither electromagnetic nor color interactions. In SUSY models which are realized with R-parity conservation the lightest supersymmetric particle (LSP) is stable. A popular cold dark matter candidate is the LSP, provided that it is electrically and color neutral. Certainly the most theoretically developed LSP is the lightest neutralino~\cite{neutralino}. However, there are other dark matter candidates as well, for example the gravitino~\cite{gravitino, steffen} and the axino~\cite{axino}, the superpartner of axion~\cite{wilczek} which solves the QCD problem via the Peccei-Quinn mechanism~\cite{quinn}. In this article we work in the framework of Randall-Sundrum type II brane model (RSII), we assume that the gravitino is the LSP and address the question whether the gravitino can play the role of dark matter in the universe, and for which range for gravitino mass and five-dimensional Planck mass. We remark in passing that gravitino production in brane cosmology has also been considered in~\cite{Mazumdar:2000ch}.

For a better qualitative understanding of the present work, we report here the main outcome of this article. The usual gravitino overproduction problem can be avoided because of the faster cosmological expansion in early RSII cosmology. Thus, the gravitino becomes a viable dark matter candidate for a wide range of parameters of the model. This requires, in particular, that the fundamental Planck mass $M_5$ is not too high. A key quantity in our analysis is the transition temperature $T_t$, which is determined solely by $M_5$. Both thermal and non-thermal gravitino production mechanisms are affected by the novel expansion law of the universe and are modified as follows. In standard four-dimensional cosmology the gravitino yield from thermal production is proportional to the reheating temperature $T_R$ after inflation. However, in RSII brane model considered here the gravitino yield becomes proportional to the transition temperature. Furthermore, the gravitino yield from non-thermal production differs from the standard expression by a factor which is inversely proportional to $T_t$.

Our work is organized as follows: The article consists of four section, of which this introduction is the first. In the second section we present the theoretical framework, while in section 3 we show the results of our analysis. Finally we conclude in the last section.

\section{The theoretical framework}

\subsection{The brane model}

Over the last years the brane-world models have been attracting a lot of
attention as a novel higher-dimensional theory. Brane models are inspired from
M/string theory and although they are not yet derivable from the fundamental
theory, at least they contain the basic ingredients, like extra dimensions,
higher-dimensional objects (branes), higher-curvature corrections to gravity
(Gauss-Bonnet) etc. Since string theory claims to give us a fundamental
description of nature it is important to study what kind of cosmology it
predicts. Furthermore, despite the fact that supersymmetric dark matter has been
analyzed in standard four-dimensional cosmology, it is challenging to discuss
it in alternative gravitational theories as well. Neutralino dark matter in brane cosmology has been studied in~\cite{okada1}, while axino dark matter in brane-world cosmology has been studied in~\cite{panotop}.

In brane-world models it
is assumed that the standard model particles are confined on a 3-brane while
gravity resides in the whole higher dimensional spacetime. The model first
proposed by Randall and Sundrum (RSII)~\cite{rs}, is a simple and interesting
one, and its cosmological evolutions have been intensively
investigated. An incomplete list can be seen e.g. in~\cite{langlois}. In the present
work we would like to study
gravitino dark matter in the framework of RSII model. According to that
model, our 4-dimensional
universe is realized on the 3-brane with a positive tension located at the UV
boundary of 5-dimensional AdS spacetime. In the bulk there is just a
cosmological constant $\Lambda_{5}$, whereas on the brane there is matter with
energy-momentum tensor $\tau_{\mu \nu}$. Also, the five dimensional Planck
mass is denoted by $M_{5}$ and the brane tension is denoted by $T$.

If Einstein's equations hold in the five dimensional bulk, then it has been shown in \cite{shiromizu} that the effective four-dimensional Einstein's equations induced on the brane can be written as
\begin{equation}
G_{\mu \nu}+\Lambda_{4} g_{\mu \nu}=\frac{8 \pi}{m_{pl}^2} \tau_{\mu \nu}+(\frac{1}{M_{5}^3})^2 \pi_{\mu \nu}-E_{\mu \nu}
\end{equation}
where $g_{\mu \nu}$ is the induced metric on the brane, $\pi_{\mu \nu}=\frac{1}{12} \: \tau \: \tau_{\mu \nu}+\frac{1}{8} \: g_{\mu \nu} \: \tau_{\alpha \beta} \: \tau^{\alpha \beta}-\frac{1}{4} \: \tau_{\mu \alpha} \: \tau_{\nu}^{\alpha}-\frac{1}{24} \: \tau^2 \: g_{\mu \nu}$, $\Lambda_{4}$ is the effective four-dimensional cosmological constant, $m_{pl}$ is the usual four-dimensional Planck mass and $E_{\mu \nu} \equiv C_{\beta \rho \sigma} ^\alpha \: n_{\alpha} \: n^{\rho} \: g_{\mu} ^{\beta} \: g_{\nu} ^{\sigma}$ is a projection of the five-dimensional Weyl tensor $C_{\alpha \beta \rho \sigma}$, where $n^{\alpha}$ is the unit vector normal to the brane.
The tensors $\pi_{\mu \nu}$ and $E_{\mu \nu}$ describe the influence of the bulk in brane dynamics. The five-dimensional quantities are related to the corresponding four-dimensional ones through the relations
\begin{equation}
m_{pl}=4 \: \sqrt{\frac{3 \pi}{T}} \: M_{5}^3
\end{equation}
and
\begin{equation}
\Lambda_{4}=\frac{1}{2 M_{5}^3} \left( \Lambda_{5}+\frac{T^2}{6 M_{5}^3} \right )
\end{equation}
In a cosmological model in which the induced metric on the brane $g_{\mu \nu}$ has the form of  a spatially flat Friedmann-Robertson-Walker model, with scale factor $a(t)$, the Friedmann-like equation on the brane has the generalized form~\cite{langlois}
\begin{equation}
H^2=\frac{\Lambda_{4}}{3}+\frac{8 \pi}{3 m_{pl}^2}  \rho+\frac{1}{36 M_{5}^6} \rho^2+\frac{C}{a^4}
\end{equation}
where $C$ is an integration constant arising from $E_{\mu \nu}$. The cosmological constant term and the term linear in $\rho$ are familiar from the four-dimensional conventional cosmology. The extra terms, i.e the ``dark radiation'' term and the term quadratic in $\rho$, are there because of the presence of the extra dimension. Adopting the Randall-Sundrum fine-tuning
\begin{equation}
\Lambda_{5}=-\frac{T^2}{6 M_{5}^3}
\end{equation}
the four-dimensional cosmological constant vanishes. In addition, the dark radiation term is severely constrained by the success of the Big Bang Nucleosynthesis (BBN), since the term behaves like an additional radiation at the BBN era \cite{orito}. So, for simplicity, we neglect the term in the following analysis. The five-dimensional Planck mass is also constrained by the BBN, which is roughly estimated as $M_{5} \geq 10 \: TeV$ \cite{cline}. The generalized Friedmann equation takes the final form
\begin{equation}
H^2=\frac{8 \pi G}{3} \rho \left (1+\frac{\rho}{\rho_0} \right )
\end{equation}
where
\begin{equation}
\rho_0=96 \pi G M_{5}^6
\end{equation}
with $G$ the Newton's constant. One can see that the evolution of the early universe can be divided into two eras. In the low-energy regime $\rho \ll \rho_0$ the first term dominates and we recover the usual Friedmann equation of the conventional four-dimensional cosmology. In the high-energy regime $\rho_0 \ll \rho$ the second term dominates and we get an unconventional expansion law for the universe. In between there is a transition temperature $T_t$ for which $\rho(T_t)=\rho_0$. Once $M_{5}$ is given, the transition temperature $T_{t}$ is determined as
\begin{equation}
T_{t}=1.6 \times 10^{7} \: \left ( \frac{100}{g_{eff}} \right )^{1/4} \: \left ( \frac{M_{5}}{10^{11} \: GeV} \right )^{3/2} \: GeV
\end{equation}
where $g_{eff}$ counts the total number of relativistic degrees of freedom.
\subsection{The particle physics model}

The extension of standard model (SM) of particle physics based on SUSY is the minimal supersymmetric standard model (MSSM)~\cite{mssm}. It is a supersymmetric gauge theory based on the SM gauge group with the usual representations (singlets, doublets, triplets) and on $\mathcal{N}=1$ SUSY. Excluding gravity, the massless representations of the SUSY algebra are a chiral and a vector supermultiplet. The gauge bosons and the gauginos are members of the vector supermultiplet, while the matter fields (quarks, leptons, Higgs) and their superpartners are members of the chiral supermultiplet. The Higgs sector in the MSSM is enhanced compared to the SM case. There are now two Higgs doublets, $H_u, H_d$, for anomaly cancelation requirement and for giving masses to both up and down quarks. After electroweak symmetry breaking we are left with five physical Higgs bosons, two charged $H^{\pm}$ and three neutral $A,H,h$ ($h$ being the lightest). Since we have not seen any superpartners yet SUSY has to be broken. In MSSM, SUSY is softly broken by adding to the Lagrangian terms of the form
\begin{itemize}
\item Mass terms for the gauginos $\tilde{g}_i$, $M_1, M_2, M_3$
\begin{equation}
M \tilde{g} \tilde{g}
\end{equation}
\item Mass terms for sfermions $\tilde{f}$
\begin{equation}
m_{\tilde{f}}^2 \tilde{f}^{\dag} \tilde{f}
\end{equation}
\item Masses and bilinear terms for the Higgs bosons $H_u, H_d$
\begin{equation}
m_{H_u}^2 H_u^{\dag} H_u+m_{H_d}^2 H_d^{\dag} H_d+B \mu (H_u H_d + h.c.)
\end{equation}
\item Trilinear couplings between sfermions and Higgs bosons
\begin{equation}
A Y \tilde{f}_1 H \tilde{f}_2
\end{equation}
\end{itemize}
In the unconstrained MSSM there is a huge number of unknown parameters~\cite{parameters} and thus little predictive power. However, the Constrained MSSM (CMSSM) or mSUGRA~\cite{msugra} is a framework with a small controllable number of parameters, and thus with much more predictive power. In the CMSSM there are four parameters, $m_0, m_{1/2}, A_0, tan \beta$, which are explained below, plus the sign of the $\mu$ parameter from the Higgs sector. The magnitude of $\mu$ is determined by the requirement for a proper electroweak symmetry breaking, its sign however remains undetermined. We now give the explanation for the other four parameters of the CMSSM
\begin{itemize}
\item Universal gaugino masses
\begin{equation}
M_1(M_{GUT})=M_2(M_{GUT})=M_3(M_{GUT})=m_{1/2}
\end{equation}
\item Universal scalar masses
\begin{equation}
m_{\tilde{f}_i}(M_{GUT})=m_0
\end{equation}
\item Universal trilinear couplings
\begin{equation}
A_{i j}^u(M_{GUT}) = A_{i j}^d(M_{GUT}) = A_{i j}^l(M_{GUT}) = A_0 \delta_{i j}
\end{equation}
\item
\begin{equation}
tan \beta \equiv \frac{v_1}{v_2}
\end{equation}
where $v_1, v_2$ are the vevs of the Higgs doublets and $M_{GUT} \sim 10^{16}~GeV$ is the Grand Unification scale.
\end{itemize}

\section{Analysis and results}

We consider four benchmark models (shown in the table below) for natural values of $m_0, m_{1/2}$, representative values of $tan \beta$ and fixed $A_0=0, \mu >0$.
In these models the lightest neutralino (denoted by $\chi$) is the lightest of the usual superpartners and thus the NLSP.
Furthermore the following experimental constraints (for Higgs, scalar $\tau$ and chargino masses and a rare decay)~\cite{steffen, precision} are satisfied

\begin{eqnarray}
m_h & > & 114.4~GeV \\
m_{\tilde{\tau}_1} & > & 81.9~GeV \\
m_{\tilde{\chi}_1^{\pm}} & > & 81.9~GeV \\
BR(b \rightarrow s \gamma) & = & (3.39_{-0.27}^{+0.30}) \times 10^{-4}
\end{eqnarray}

\begin{table}[ht]
\begin{center}
\begin{tabular}{|c|c|c|c|c|c|}
\hline
 Model & $m_0 \: (GeV)$ & $m_{1/2} \: (GeV)$ & $tan \beta$ & $m_{\chi} \: (GeV)$ & $\Omega_{\chi} h^2$  \\ \hline
 A & 200 & 500 & 15 & 205.42 & 0.64 \\ \hline
 B & 400 & 800 & 25 & 337.95 & 1.82 \\\hline
 C & 1000 & 600 & 30 & 252.41 & 7.37 \\ \hline
 D & 350 & 450 & 20 & 184.46 & 1.2 \\ \hline
\multicolumn{6}{l}{Table 1: The four benchmark models considered in the analysis}
\end{tabular}
\end{center}
\end{table}

The SUSY spectrum (as well as the Higgs bosons masses) is computed using the FORTRAN code SuSpect~\cite{suspect}, the neutralino relic density is computed using the computer program micrOMEGAs~\cite{micromegas} and the top quark mass is fixed to $m_t=172.7~GeV$~\cite{cdf}.

For the gravitino abundance we take into account both thermal (TP) and non-thermal production (NTP)
\begin{equation}
\Omega_{3/2} h^2 = \Omega_{3/2}^{NTP} h^2 + \Omega_{3/2}^{TP} h^2
\end{equation}
and we impose the WMAP constraint for cold dark matter~\cite{wmap}
\begin{equation}
0.075 < \Omega_{cdm} h^2=\Omega_{3/2} h^2 < 0.126
\end{equation}
In the NTP case the contribution to the gravitino abundance comes from the decay of the NLSP
\begin{equation}
\Omega_{3/2} h^2 = \frac{m_{3/2}}{m_{NLSP}} \: \Omega_{NLSP} h^2
\end{equation}
with $m_{3/2}$ the gravitino mass, $m_{NLSP}$ the mass of the NLSP and $\Omega_{NLSP} h^2$ the NLSP abundance had it did not decay into the gravitino.
In the TP case the gravitino abundance is given by
\begin{equation}
\Omega_{3/2} h^2=\frac{m_{3/2} s(T_0) Y_{3/2}^{TP} h^2}{\rho_{cr}}=2.75 \times 10^{8} \left ( \frac{m_{3/2}}{GeV} \right ) Y_{3/2}^{TP}(T_0)
\end{equation}
where $T_0=2.7~K$ is the temperature of the universe today, $s=h_{eff} 2 \pi^2/45$ is the entropy density for the relativistic degrees of freedom, $\rho_{cr}=(3 H_0^2)/(8 \pi G)$ is the critical energy density today and $Y_{3/2}$ is the gravitino yield defined by $Y_{3/2} \equiv n_{3/2}/s$, where $n_{3/2}$ is the gravitino number density. For the yield in standard four-dimensional cosmology we employ the formula (3) of~\cite{steffen}, which updates~\cite{Pradler:2006qh} an older calculation~\cite{bolz}. We see that $Y_{3/2}^{TP}(T_0)$ is proportional to the reheating temperature after inflation.

Now we need to take into account the effects of the novel law for expansion of the universe. These are twofold: First, in the TP case the reheating temperature is replaced by the transition temperature, $T_R \rightarrow 2T_t$~\cite{panotop, okada3}. Second, the relic density of a particle of mass $m$ is modified as follows~\cite{okada2}
\begin{equation}
\frac{\Omega^{(b)}}{\Omega^{(s)}}=C \: \frac{x_t}{x_{d}^{(s)}}
\end{equation}
where C is a numerical factor of order one, the index b stands for "brane", the index s stands for "standard", $x_t=m/T_t$ and $x_d=m/T_d$, with $T_t$ the transition temperature and $T_d$ the decoupling temperature of the particle of mass $m$. In standard cosmology $x_d^{(s)} \simeq 30$.

For each benchmark model we have obtained plots (below we show as an example the plots for model A, for the rest of the models there are similar plots) which show the gravitino abundance $\Omega_{3/2} h^2$ as a function of the gravitino mass $m_{3/2}$ for several values of the five-dimensional Planck mass $M_5$. We discriminate between low values and high values of $M_5$. In the TP case, the calculation assumes that the particles are in thermal equilibrium and thus the analysis is valid for temperatures $T > 1~TeV$ or for five-dimensional Planck mass $M_5 > 1.83 \times 10^8~GeV$. We use this criterion to divide the values of $M_5$ into two categories, low values $M_5 < 1.83 \times 10^8~GeV$ and high values $M_5 > 1.83 \times 10^8~GeV$. Figure $1$ corresponds to the first case (the values that we have used are $5 \times 10^4~GeV$, $8 \times 10^5~GeV$ and $6 \times 10^6~GeV$ from top to bottom), while figure $2$ corresponds to the second case (the values that we have used are $10^{10}~GeV$, $10^{11}~GeV$ and $10^{12}~GeV$ from bottom to top). In the low $M_5$ case there is always one allowed range for the gravitino mass from the sub-$GeV$ range to a few tens of $GeV$. On the other hand, in the high $M_5$ case there are two allowed ranges for the gravitino mass, one in the sub-$GeV$ range and another one in the -a few $GeV$ up to a few tens of $GeV$- range. If however $M_5$ is high enough, $M_5 > 10^{12}~GeV$, there is no allowed range at all.

\section{Conclusions}

We have studied gravitino dark matter in the brane-world cosmology. The theoretical framework for our work is the CMSSM for particle physics and RS II for gravity, which predicts a generalized Friedmann-like equation for the evolution of the universe. We assume that gravitino is the LSP and the lightest neutralino is the NLSP. For the gravitino abundance we have taken into account both thermal and non-thermal production and have imposed the cold dark matter constraint $0.075 < \Omega_{cdm} h^2 < 0.126$. The formulae valid in standard four-dimensional cosmology are corrected taking into account the novel expansion law for the universe. We have considered four benchmark models for natural values of $m_0$ and $m_{1/2}$ and representative values of $tan \beta$. In these models the neutralino is the lightest of the usual superpartners (and thus the NLSP, since we assume that the gravitino is the LSP) and experimental constraints are satisfied. For each benchmark model we have produced plots of the gravitino abundance as a function of the gravitino mass for several different values of the five-dimensional Planck mass. The obtained plots show that in general the gravitino can be the cold dark matter in the universe for gravitino masses from a few $GeV$ up to tens of $GeV$. Furthermore an upper bound on the five-dimensional Planck mass is obtained, $M_5 < 10^{12}~GeV$.

\section*{Acknowlegements}

We would like to thank T.~N.~Tomaras for a careful reading of the manuscript. This work was supported by project "Particle Cosmology".

\newpage

\begin{figure}
\centerline{\epsfig{figure=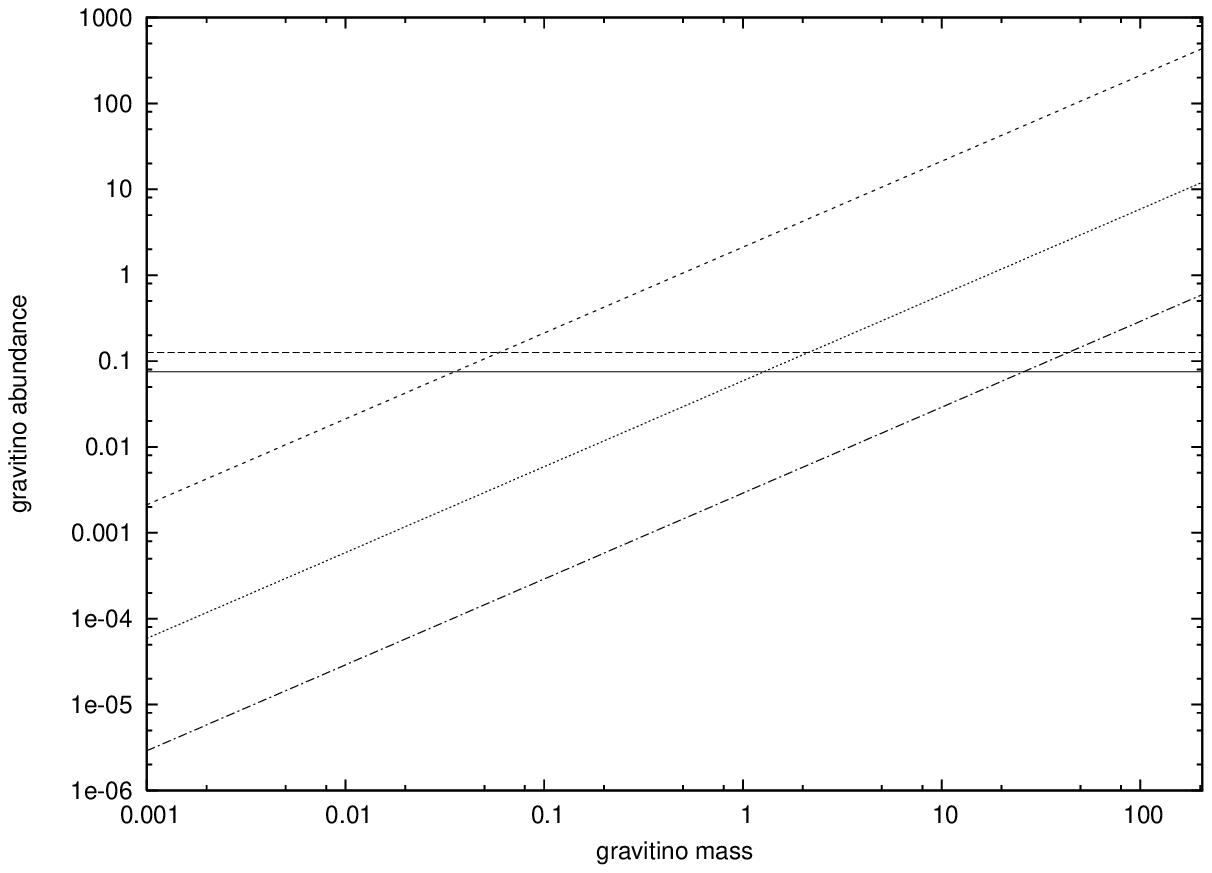,height=8cm,angle=0}}
\caption{Gravitino abundance versus gravitino mass (in $GeV$) for several values of the five-dimensional Planck mass (benchmark model A and low $M_5$). The strip around $0.1$ is the allowed range for cold dark matter. Values of $M_5$ used are $5 \times 10^4~GeV$, $8 \times 10^5~GeV$ and $6 \times 10^6~GeV$ from top to bottom.}
\end{figure}

\begin{figure}
\centerline{\epsfig{figure=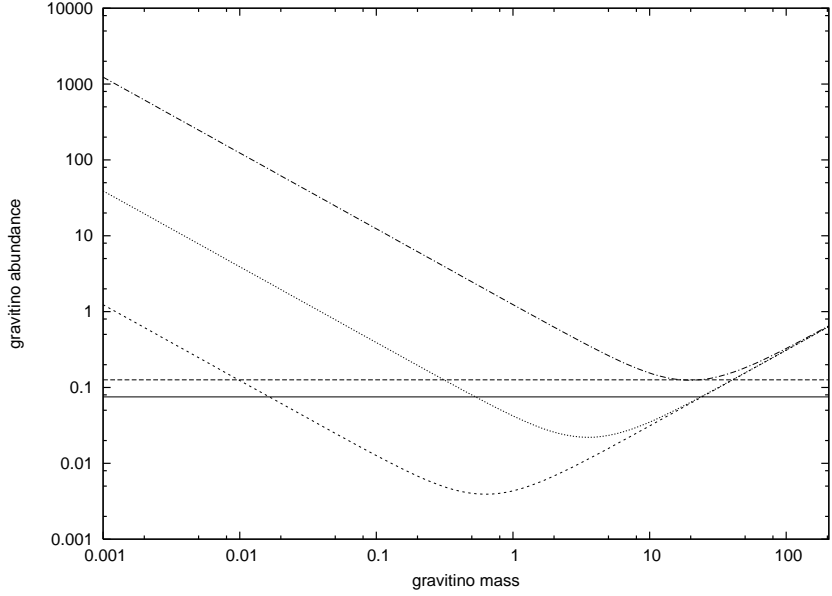,height=8cm,angle=0}}
\caption{Same as figure $1$, but for high $M_5$. Values of $M_5$ used are $10^{10}~GeV$, $10^{11}~GeV$ and $10^{12}~GeV$ from bottom to top.}
\end{figure}

\end{document}